\documentclass[preprint,showpacs,preprintnumbers,amsmath,amssymb]{revtex4}
\usepackage{graphicx}
\usepackage{dcolumn}
\usepackage{bm}

\begin{document}
\def\bbox#1{\hbox{\boldmath${#1}$}}

\title[]{Relativistic Generalization of the Gamow Factor\\ for Fermion Pair Production
or Annihilation}
 
\author{Jin-Hee Yoon}
\email{jinyoon@inha.ac.kr}
\affiliation{Department of Physics, Inha University, Inchon, South Korea}
\author{Cheuk-Yin Wong}
\email{wongc@ornl.gov}
\affiliation{
Physics Division, Oak Ridge National Laboratory, Oak Ridge, TN
37831, U.S.A.\\
Department of Physics, University of Tennessee, Knoxville, TN
37996, U.S.A.
}

\received{\today}

\begin{abstract}
In the production or annihilation of a pair of fermions, the
initial-state or final-state interactions often lead to significant
effects on the reaction cross sections.  For Coulomb-type
interactions, the Gamow factor has been traditionally used to take
into account these effects.  However the Gamow factor needs to be
modified when the magnitude of the coupling constant or the relative
velocity of two particles increases. We obtain the relativistic
generalization of the Gamow factor in terms of the overlap of the
Feynman amplitude with the relativistic wave function of two fermions
with an attractive Coulomb-type interaction.  An explicit form of the
corrective factor is presented for the spin-singlet S-wave state.
While the corrective factor approaches the Gamow factor in the
non-relativistic limit, we found that the Gamow factor significantly
over-estimates the effects when the coupling constant or the velocity
is large.
\end{abstract}


\maketitle

\narrowtext

\section{Introduction} 

The final-state interaction (FSI) and initial-state interaction (ISI)
are important processes in particle and nuclear physics
\cite{Bet56,Gam28,Som39,Bar80,Gus88,Fad88,Fad90,Bro95,Cha95a,Won96d,Won97,Yoon00,Yoon03}.
They lead to an enhancement of the reaction cross section for
attractive interactions and a suppression for repulsive interactions.
The effects are especially large near the production threshold
in the region of low-energy annihilation.  We shall describe these
FSI/ISI in terms of a corrective factor which we shall
call the $K$-factor.  It is defined as the ratio of the cross section
with the interaction to the corresponding quantity without the
interaction.

In non-relativistic physics the $K$-factor can be obtained by solving
the two-body Schr\"odinger equation nonperturbatively under their
mutual interaction.  It is determined by calculating the absolute
square of the relative wave function at the origin.  As is well known,
for the electric-Coulomb and color-Coulomb interaction, $V(r)=-\alpha/r$, the non-relativistic corrective $K$-factor is the
Gamow-Sommerfeld factor \cite{Gam28,Som39} (or simply called, the
Gamow factor),
\begin{eqnarray}
\label{eq:80}
G (\eta)
= 
{ 2 \pi \eta \over 1- e^{-2 \pi \eta}  },
\end{eqnarray}
where $\eta$ is the Sommerfeld parameter, 
\begin{eqnarray}
\label{eq:xi}
\eta={ \alpha \over v },
\end{eqnarray}
$\alpha$ is the coupling constant (positive for an attractive Coulombic
interaction), and $v$ is the relative velocity of
two particles.  For two-equal masses which we shall consider in
this paper, $v$ is given in terms of their center-of-mass energy
$\sqrt{s}$ by \cite{Yoon00,Tod71,Cra91},
\begin{eqnarray}
\label{eq:v}
v= {(s^2 - 4s m^2)^{1/2} \over s - 2 m^2 }.
\end{eqnarray}
This gives $v \sim 2\sqrt{1-4m^2/s}$ when $\sqrt{s} \sim 2m$ and
$v\rightarrow 1$ when $s \rightarrow \infty$.  

As was pointed out by Chatterjee and her collaborator
\cite{Cha95a,Won96d,Won97}, the Gamow factor Eq.\ (\ref{eq:80}), has
been traditionally used to study non-relativistic Coulomb-type ISI/FSI
in reactions involving production or annihilation of particles.  The
results are then interpolated with the well-known perturbative QCD
corrective $K$-factor at high energies, following the procedure of
Schwinger \cite{Sch73}.  It predicts an enhancement for $q\bar q$ in
color-singlet states and a suppression for color-octet states, the
effect increasing as the relative velocity decreases. Consequences on
dilepton production in the quark-gluon plasma, the Drell-Yan process,
and heavy quark production processes were also examined.

Although the above Gamow factor gives an approximate description of
the FSI/ISI effects \cite{Cha95a}, it is useful to obtain a more
accurate description as there are physical processes in which the
interaction coupling constant or the relative velocity of the pair can
be quite large and the use of the non-relativistic treatment may not
be adequate.  For example, in the production of a charm quark pair,
the coupling constant of the color-Coulomb interaction between the
charm quark and antiquark is about $0.3$, which is quite
large. Furthermore, as the charm quark mass is large, the magnitude of
the relative velocity between the produced quark and antiquark can be
quite small in low-energy production near the threshold.  The FSI/ISI
effects can be quite large for large coupling constants and small
relative velocities.  Baym and P. Braun-Munzinger modified the Gamow
factor in their study of the final-state Coulomb interaction and
effects on the Hanbury-Brown Twiss effects of intensity interferometry
\cite{Bay96}.  A negatively charged particle in a nucleus with a large
$Z$ number will also be subject to strong FSI/ISI.  Although the
effect of the interaction is very large for low relative velocities,
it is nonetheless useful to see how the effect varies as the velocity
increases.  For brevity of notation, we shall use the term ``Coulomb
interaction'' with a variable coupling constant to refer to both the
electric-Coulomb and color-Coulomb interactions.  We shall limit our
attention to attractive Coulomb interactions, although similar
formulation can be carried out for repulsive Coulomb interactions and
screened Yukawa interactions \cite{Cha95a,Won96d,Won97}.

Relativistic treatment is needed for strongly attractive interactions,
even when the relative asymptotic velocity between two particles
at $r\to \infty$ is small, as two particles can reach relativistic
velocities at short distances due to the strongly attractive
interaction.  Relativistic treatment is also needed when the asymptotic
relative velocity of two particle approaches the speed of light.

The evaluation of the relativistic corrective $K$-factor involves the
non-perturbative treatment of the relativistic two-body equation of
motion under their mutual interaction.  Compared with the
non-relativistic Schr\" odinger equation involving the Coulomb
potential, there is an additional attractive effective potential
proportional to $-|V(r)|^2$, and a repulsive term from the
space-like part of the gauge interaction, which lead to a non-trivial
behavior when the coupling constant becomes large.  In the case of
fermions with the Coulomb interaction, there are further modifications
associated with additional spin-dependent potentials.

In a previous study~\cite{Yoon00} we presented a method to study the
relativistic Coulomb FSI/ISI effects for a pair of bosons.  The
corrective $K$-factor was evaluated by taking the overlap of the
relativistic wave function with the corresponding Feynman
amplitude. Its analytic form was obtained and its numerical values
were compared with those of the Gamow factor. For attractive
interactions, we found that the Gamow factor over-estimates the
corrective factor for most energies and even more in the
relativistic region with a large magnitude of the coupling
constants. 

We would like to generalize these previous boson results to a pair of
fermions with an attractive Coulomb-type final state interaction.  
The fermion results are of more practical interest as the
electromagnetic gauge field or the chromodynamical gauge field couples
to fermion fields and the results obtained here may be directly
applied to the production or annihilation of a pair of fermions 
in a color-singlet state.
A brief report of the present results was presented in ref. \cite{Yoon03}

\section{The $K$-factor}

We shall be interested in processes involving the production or
annihilation of a pair of fermions $p_1$ and $p_2$ with equal masses
by photons (or gluons) of momenta $k_1$ and $k_2$ represented by the
Feynman diagrams in Fig.\ 1.  In these diagrams, a solid line
represents a fermion and a wavy line can be either a photon or a
gluon.  We shall evaluate the $K$-factor using a boson-fermion
interaction vertex coupling constant $g$, which will be canceled
out in the ratio in Eq.\ (\ref{Kfac}) below.  The $K$-factor depends
on the FSI/ISI between the fermions and does not depend on how the
pair of fermions is produced or annihilated.  For definiteness, we
shall study the production process $k_1 + k_2 \rightarrow p_1 + p_2$.

\begin{figure}[t]
\includegraphics[scale=0.8]{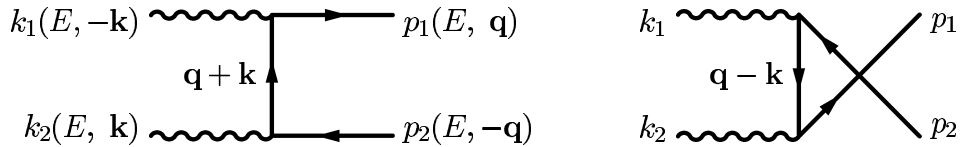}
\vspace*{-18.3cm}\hspace*{1cm}
\caption{The Feynman diagrams included in the calculation
  \label{fig:feynman}}
\end{figure}

The simplest description of the process is to assume that there is no
FSI/ISI and the probability amplitude for the production of this
pair of particles $p_1$ and $p_2$ can be determined by means of
perturbation theory.  The state $|\Phi_{p_1 p_2}\rangle$ of the $p_1 p_2$ pair
after the reaction $k_1 + k_2 \rightarrow p_1 + p_2$ is represented by
the state vector
\begin{eqnarray}
\label{eq:Phi}
|\Phi_{p_1 p_2} \rangle =
{\cal M}(k_1 k_2
\rightarrow p_1 p_2 )| p_1 p_2 \rangle,
\end{eqnarray}
where ${\cal M}(k_1k_2 \rightarrow p_1p_2)$ is the Feynman amplitude for the
process.  For two-particle system
$p_1p_2$, we define the center-of-mass momentum ${\bm P}={\bm p_1}+{\bm p_2}$ and the
relative momentum ${\bm q}=({\bm p_1}-{\bm p_2})/2$.
Therefore, ${\bm p_1} = {\bm P}/2 + {\bm q}$ and ${\bm p_2} = {\bm P}/2 - {\bm q}$.

On the other hand, under their mutual interaction between $p_1$ and
$p_2$, we can describe an interacting $p_1p_2$ pair with a
center-of-mass momentum ${\bm P}$ as
\begin{eqnarray}
\label{eq:Psi}
|\Psi_V \rangle= {\tilde \psi}({\bf q})|{\bm P}\rangle.
\end{eqnarray}

We introduce the corrective $K$-factor defined as the ratio of the
cross sections with and without the FSI/ISI
\cite{Pes95,Cra91,Won97,Yoon00}
\begin{eqnarray}
\label{Kfac}
K = {\sigma_V \over \sigma_0} =
{ |<\Psi_V|\Phi_{p_1 p_2}>|^2 \over |<\Psi_0|\Phi_{p_1 p_2}>|^2 } 
\end{eqnarray}
where the scalar product $\langle \Psi_V|\Phi_{p_1 p_2}\rangle$ gives
the probability amplitude for the produced pair 
$|\Phi_{p_1 p_2}\rangle$ to be in the interacting state $|\Psi_V
\rangle$,
\begin{eqnarray}
\label{eq:amp}
\langle\Psi_V |\Phi_{p_1 p_2}\rangle = \int {d{\bf q} \over (2\pi)^3}
{\tilde \psi}^{\dagger}({\bf q}) {\cal M}(k_1 k_2 \rightarrow 
p_1({\bf q}) p_2({\bf -q}))
\end{eqnarray}
and the scalar product $\langle\Psi_0 |\Phi_{p_1 p_2}\rangle$ is
similarly defined in terms of the wave function ${\tilde{\psi}}_0{(
{\bf q} )}$ for a pair of free fermions.  The corrective $K$-factor
should approach the Gamow factor in the non-relativistic limit.

The cross section with the FSI/ISI corrections is obtained simply by
multiplying the lowest order cross section with this corrective
$K$-factor,
\begin{eqnarray}
\label{eq:rat1}
{ {\hbox{Production~or~annihilation~}}
\choose {\hbox{cross~section~with~FSI/ISI~($\sigma_V$) }} }
= K \times 
{ {\hbox{Production~or~annihilation~}}
\choose {\hbox{cross~section~without ~FSI/ISI~($\sigma_0$)}} }
\end{eqnarray}

\section{Dirac Equation for the Coulomb Interaction}

To obtain the two-body wave function, we use the relativistic two-body
equation as formulated in Dirac's constraint dynamics
\cite{Dir64,Cra82,Van86,Cra87,Cra88,Tod71,Cra91} for two fermions with
4-momenta $p_1$ and $p_2$.  We choose to work in the center-of-mass
system in which ${\bm P}=(\sqrt{s}, {\bf 0})$ and ${\bm q}=(0,{\bf
q})$.  In the absence of any interaction, we have the relation between
an effective energy $\epsilon_w$, and a generalized reduced mass $m_w$
as given by
\begin{eqnarray}
\label{eq:NI}
\epsilon_w^2-{\bf q}^2- m_w^2 = 0,
\end{eqnarray}
where
\begin{eqnarray}
\label{eq:ew}
\epsilon_w={s - 2m^2  \over 2 \sqrt{s}} 
\end{eqnarray}
and
\begin{eqnarray}
m_w={m^2 \over \sqrt{s}}.
\end{eqnarray}

Next, in the case when the two fermions interact with a mutual
Coulomb-type interaction
\begin{eqnarray}
V(r)=- {\alpha \over r},
\end{eqnarray}
the solution for the two spin-$1 \over 2$ system under a mutual
interaction $V(r)$ can be written as \cite{Van86}
\begin{equation}
\label{eq:ds}
 \psi_D = -2mE_m \left[ 
	\begin{array}{c}
	 \psi \\
	 {1 \over E_m} {\bf \sigma_1} \cdot {\bf q} ~\psi \\
	 -{1 \over E_m} {\bf \sigma_2} \cdot {\bf q} ~\psi \\
	 - {1 \over E_m^2} {\bf \sigma_1} \cdot {\bf q} 
		~ \sigma_2 \cdot {\bf q} ~\psi
	\end{array} \right]
\end{equation}
where 
\[ E_m \equiv \sqrt{s}/2 - V(r) 
	+ m \sqrt{ 1-2V(r)/\sqrt{s} }. \] 
To remove the complications brought by the spinor algebra, 
we shall carry out calculations for the production
of the singlet ($S=0$) system.
The spin-singlet state is governed by the
following equation of relative motion \cite{Van86,Cra87,Cra91}
\begin{eqnarray}
\label{eq:dirac}
\biggl \{ [\epsilon_w - V (r) ]^2 - {\bf q}^2 - m_w^2 \biggr \} 
\psi = 0.
\end{eqnarray}
By factoring off the angular dependence and the spin dependence:
$\psi({\bf r})= R_{l}(r) Y_{lm}(\theta,\phi)\chi^{(S=0)}$. 
The Schr\"odinger-like radial equation for the state can be written as
\begin{eqnarray}
\label{eq:psieq}
\biggl [ {d \over dz^2} + { 2 \over z} { d \over dz} - { l (l+1) \over
z^2 } + {2 \eta \over z} + {\alpha^2 \over z^2 } + 1 \biggr ]
R_{l}(r)= 0,
\end{eqnarray}
where $z=p r$, $p$ is the asymptotic momentum at
$r\rightarrow \infty$ given by
\begin{eqnarray}
\label{eq:pin}
p= \sqrt{\epsilon_w^2 - m_w^2}.
\end{eqnarray}
The wave function $R_{l}(r)$ can be represented by the dimensionless
variable $z=p r$ and is characterized by two dimensionless parameters:
$\eta={\alpha / v}$ and $\alpha^2$, where $v={p / \epsilon_w}$ is
given by Eq.\ (\ref{eq:v}).  The solution of Eq.\ (\ref{eq:psieq}) is
\begin{eqnarray}
\label{eq:wf}
R_{l}(r)= {\left | \Gamma (a) \right| \over \Gamma (b)} e^{\pi\eta /
2} (2iz)^{\mu - 1/2} e^{-iz} {}_1F_1 (a,b,2iz)
\end{eqnarray}
where 
\begin{eqnarray}
a= \mu + {1 \over 2} + i \eta,
\end{eqnarray}
\begin{eqnarray}
b=2\mu+1,
\end{eqnarray}
\begin{eqnarray}
\mu=\sqrt{ ( l+{ 1 \over 2})^2 - \alpha^2 },
\end{eqnarray}
and the normalization constant has been determined by using the
boundary condition that at $r \rightarrow \infty$, $R_{l}(r)
\rightarrow \sin (p r + \delta_l)/p r $ with the Coulomb phase shift
$\delta_l$.  For the singlet S-state, the critical value of $\alpha$
is 1/2.

\section{The Feynman Amplitude and the Overlap with the Coulomb Wave Function}

We consider the production of the pair of fermions from the fusion of
two gluons (or photons) as in the Feynman diagram of Fig.\ 1.  Because
the relevant factors associated with the mode of production will be
canceled out in Eq.\ (\ref{Kfac}) when we take the ratio, the results
of the $K$-factor depend only on the final-state interaction.

The diagrams in Fig.\ 1 give the amplitude
\begin{eqnarray}
-i {\cal M}(k_1 k_2 \rightarrow p_1({\bm q})p_2({\bm -q})) 
   =&& (-i) e^2 \bar{v}^{(2)}(E,-{\bf q})
     \biggl [ {\bf \gamma} \cdot {\bbox{ \epsilon}}_1 
	{ {-i{\bf \gamma} \cdot ({\bm q} + {\bm k}) + m}
	 \over {({\bm q}+{\bm k})^2+ m^2 } }
	 {\bf \gamma} \cdot {\bbox{\epsilon}}_2 
\nonumber\\
	 && ~~~~ + {\bf \gamma} \cdot {\bbox{\epsilon}}_2
	{ {-i{\bf \gamma} \cdot ({\bm q} - {\bm k}) + m}
	 \over {({\bm q}-{\bm k})^2+ m^2 } }
	 {\bf \gamma} \cdot {\bbox{\epsilon}}_1
     \biggr ] u^{(1)}(E,{\bf q})
\end{eqnarray}
where ${\bm k}$ is the four-momentum of the gluon, ${\bm q}$ is the
four-momentum of one of the fermions, and ${\bbox{\epsilon}_i}$ is the
polarization vector of the $i$-th gluon.  However for the production
of a pair of fermions under their mutual final-state interactions, we
need to project out from the Feymann amplitude the proper state
${\tilde \psi}({\bf q})$ representing the two fermions under final-state
interactions,
\begin{eqnarray}
\label{eq:amp1}
\langle{\Psi_V}|\Phi_{ab} \rangle
= \int {d{\bf q} \over (2\pi)^3}
{ {\tilde \psi} ({\bf q}) \over \sqrt{2} }
\biggl [ \left\{ -i {\cal M}(k_1 k_2 \rightarrow p_1({\bf q}) p_2(-{\bf q})) 
\right\}
- \left\{ (2) \leftrightarrow (1) \right\} \biggr ].
\end{eqnarray}
Here the factor $1/\sqrt{2}$ and the exchange term is added
for the singlet fermion state.
Using the spinor algebra and full 16 component calculation, the above equation leads to
\begin{eqnarray}
\label{eq:amp2}
\langle{\Psi_V}|\Phi_{ab} \rangle
= { ie^2 \over 2m(E+m) }
  \int {d{\bf q} \over (2\pi)^3} 
  \sqrt{2} {\tilde \psi} ({\bf q})
  \biggl [ 
	{ {\cal F}({\bf q},{\bf q}+{\bf k},{\bbox{\epsilon}}_1,
		   {\bbox{\epsilon}}_2)
	  \over ({\bm q}+{\bm k})^2+ m^2 } +
	{ {\cal F}({\bf q},{\bf q}-{\bf k},{\bbox{\epsilon}}_1,
		   {\bbox{\epsilon}}_2)
	  \over ({\bm q}-{\bm k})^2+ m^2 }
  \biggr ]
\end{eqnarray}
where
\begin{eqnarray}
{\cal F}({\bf q},{\bf p},{\bbox{\epsilon}}_1,{\bbox{\epsilon}}_2)
	 =&& -i(E+m)^2 {\bf p} \cdot 
		     ({\bbox{\epsilon}}_1 \times {\bbox{\epsilon}}_2 )
	-i{\bf q} \cdot {\bbox{\epsilon}}_1 \ 
	 {\bf p} \cdot {\bf q} \times {\bbox{\epsilon}}_2 
\nonumber\\
	&&+i{\bf q} \times {\bbox{\epsilon}}_1 \cdot  
	 {\bf p} \ {\bf q} \cdot {\bbox{\epsilon}}_2
	-i{\bf q}\cdot{\bf p} \ 
		{\bf q}\cdot {\bbox{\epsilon}}_1 \times {\bbox{\epsilon}}_2 .
\end{eqnarray}

As the Coulomb wave function of Eq.\ (\ref{eq:wf}) is given in the
configuration space, it is useful to write the above integral in terms
of the wave function in configuration space. The latter is given by
\begin{eqnarray}
\label{eq:four}
\psi({\bf r}) = \int { d{\bf q} \over (2\pi)^{3} } 
{\tilde \psi} ({\bf q}) e^{-i {\bf q}\cdot {\bf r}}.
\end{eqnarray}

In conventional applications, one expands the Eq. (\ref{eq:amp2}) 
in powers of $q$ and keeps only the lowest order
$q$-independent term ${\cal M}_0$:
\begin{eqnarray}
{\cal M} \approx {\cal M}_0 + O(|{\bf q}|).
\end{eqnarray}
In this approximation, Eqs.\
(\ref{eq:amp2}) and (\ref{eq:four}) give the usual $K$-factor as
the absolute square of the wave function $\psi({\bf r})$ at the origin,
\begin{eqnarray}
K=|\psi({\bf r}=0)|^2 .
\end{eqnarray}
However, such an approximation cannot be applied to our case with the
relativistic wave function since the wave function, Eq.\ (\ref{eq:wf}), is
infinite at the origin.  To avoid this singular behavior, the full
Feynman amplitude is needed to evaluate the overlap integral and the
$K$-factor in Eqs.\ (\ref{eq:amp2}) and (\ref{Kfac}). 

In terms of the spatial wave function, the overlap integral
(\ref{eq:amp2}) is
\begin{eqnarray}
\langle{\Psi_V}|\Phi_{ab} \rangle
&=& {\sqrt{2} e^2 \over m(E+m)}
  \int d{\bf r} e^{-i{\bf k}\cdot {\bf r}}
  \Biggl [ (E+m)^2 \psi({\bf r}) 
	{\nabla \over i} \left( {e^{-mr} \over 4 \pi r } \right)
	\cdot ( {\bbox{\epsilon}}_1 \times {\bbox{\epsilon}}_2 ) 
\nonumber\\
	&& + {\nabla \over i} \left( {e^{-mr} \over 4 \pi r } \right) \cdot
	\biggl \{ ({\bbox{\epsilon}}_2 \times \nabla) 
		{\bbox{\epsilon}}_1 \cdot \nabla
		- ({\bbox{\epsilon}}_1 \times \nabla) 
		{\bbox{\epsilon}}_2 \cdot \nabla 
	- \nabla ( {\bbox{\epsilon}}_1 \times {\bbox{\epsilon}}_2 )
		\cdot \nabla \biggr \} \psi({\bf r}) 
\Biggr ].
\end{eqnarray}
We shall specialize to the S-wave case with $l=0$ in the present
manuscript.  Higher partial waves can be considered in future work.
In this simple case with $l=0$, the above integral becomes
\begin{eqnarray}
\langle{\Psi_V}|\Phi_{ab} \rangle
&=& {\sqrt{2} e^2 \over m(E+m)}
  \int r^2 dr j_1 (kr) 
  \Biggl \{ -(E+m)^2 \psi(r) + {2\psi'(r) \over r} + \psi''(r)
  \biggr \} 
\nonumber\\
&& {d\over dr}\left( {e^{-mr} \over r} \right)
	 {{\bf k} \over k} \cdot ( {\bbox{\epsilon}}_1 \times {\bbox{\epsilon}}_2 ).
\end{eqnarray}

Using the differential equation (\ref{eq:psieq}) and the wave
function of Eq.\ (\ref{eq:wf}), we carry out the above integration 
and obtain
\begin{eqnarray}
\label{eq:mfi}
\langle{\Psi_V}|\Phi_{ab} \rangle
&=& {\sqrt{2} e^2 \over m(E+m)} 
  {{\bf k} \over k} \cdot ( {\bbox{\epsilon}}_1 \times {\bbox{\epsilon}}_2 )
  {k \over 3}{|\Gamma(a)| \over \Gamma(b)}
  e^{\pi \eta /2}
  \sum_{n=0}^\infty
  { (a)_n  \Gamma ({3\over 2}+\mu + n) \over (b)_n~ n! }
  \left ({  2i p \over \sqrt{\delta^2 + k^2}} \right )^{n+\mu-{1 \over2} }
\nonumber \\
&\times& \Biggl [ 
  { m \{(E+m)^2 + \epsilon_w^2 -m_w^2 \}
	\over \left(\sqrt{\delta^2 +k^2}\right)^3}
    \left({3 \over 2}+\mu+n\right)
  F({5 \over 4} + {\mu+n \over 2}, {3\over 4} - {\mu + n \over 2}; 
    {5\over 2} ; \xi^2)
\nonumber \\
  &&+ { (E+m)^2 + \epsilon_w^2 -m_w^2 + 2\epsilon_w \alpha m 
		\over \left(\sqrt{\delta^2 +k^2}\right)^2 }
  F({3 \over 4} + {\mu+n \over 2}, {5\over 4} - {\mu + n \over 2}; 
    {5\over 2} ; \xi^2)
\nonumber \\
 & &+ { \{ 2\alpha \epsilon_w + m\alpha^2 \} 
	 	\over \sqrt{\delta^2 +k^2}}
   {1 \over \left({1 \over 2}+\mu+n\right)}
  F({1 \over 4} + {\mu+n \over 2}, {7\over 4} - {\mu + n \over 2}; 
    {5\over 2} ; \xi^2)
\nonumber \\
  &&+ \alpha^2 {1 \over \left({1 \over 2}+\mu+n\right)
			\left(-{1 \over 2}+\mu+n\right)}
  F(-{1 \over 4} + {\mu+n \over 2}, {9\over 4} - {\mu + n \over 2}; 
    {5\over 2} ; \xi^2)
\biggr ]
\end{eqnarray}
where 
\begin{eqnarray}
\delta=m+ip,
\end{eqnarray}
\begin{eqnarray}
\xi^2={k^2\over (m+ip)^2 + k^2}.
\end{eqnarray}
In Eq.\ (\ref{eq:mfi}), $(a)_n=a (a+1)(a+2)..(a+n-1)$ with $(a)_0=1$.
The quantity $(b)_n$ is similarly defined.  This expression looks
similar to that in our previous results for boson case in
Ref.\ \cite{Yoon00}.

\section{Results for the $K$-factor }

We introduce the complex angle variable
\begin{eqnarray}
\theta = \tan^{-1}{ k \over m+ip} = {\pi \over 4} - i {1 \over 4}
\ln { k + p \over k - p},
\end{eqnarray}
which is a relativistic measure of the relative motion between
particles $p_1$ and $p_2$.  The real part of $\theta$ is always $\pi/4$,
and the imaginary part is negative, with a magnitude that is half of
the rapidity of $p_1$(or $p_2$) in the center-of-mass system.

In terms of the angle variable, Eq.(\ref{eq:mfi}) can be transformed
as follows:
\begin{eqnarray}
\label{eq:up}
\langle{\Psi_V}|\Phi_{ab} \rangle
= {\sqrt{2} e^2 \over m(E+m)} 
  {\bf k} \cdot ( {\bbox{\epsilon}}_1 \times {\bbox{\epsilon}}_2 )
	{|\Gamma(a)| \over \Gamma(b)}
	e^{\pi \eta /2}	 {\cal A},
\end{eqnarray}
where the factor $\cal A$ is
\begin{eqnarray}
\label{eq:A1}
{\cal A}&=& 
\sum_{n=0}^\infty
{ (a)_n   \Gamma (2\nu) \over (b)_n~ n! } 
{ \left ( {2ip \over \sqrt{\delta^2 + k^2} } \right ) ^{n+\mu-{1\over 2}} } 
\nonumber\\
&\times& \Biggl [ 
  {m\over k} {(E+m)^2 + \epsilon_w^2 -m_w^2 \over k^2}
  {2\nu \over 2\nu-2} 
  \left\{ {\sin (2\nu-1) \theta \over 2\nu-1} \cos \theta
 	- {\sin (2\nu) \theta \over 2\nu} \right\}
\nonumber\\
& &+ {(E+m)^2 + \epsilon_w^2 -m_w^2 +2\alpha m \epsilon_w \over k^2}
  {1 \over 2\nu-3} {1 \over \sin \theta}
  \left\{ {\sin (2\nu-2) \theta \over 2\nu-2} \cos \theta
 	- {\sin (2\nu-1) \theta \over 2\nu-1} \right\}
\nonumber\\
&&+  { 2\alpha \epsilon_w +m\alpha^2 \over k}
  {1 \over (2\nu-1)(2\nu-4)} {1 \over \sin^2 \theta}
  \left\{ {\sin (2\nu-3) \theta \over 2\nu-3} \cos \theta
 	- {\sin (2\nu-2) \theta \over 2\nu-2} \right\}
\nonumber\\
&&+  \alpha^2
  {1 \over (2\nu-1)(2\nu-2)(2\nu-5)} {1 \over \sin^3 \theta}
  \left\{ {\sin (2\nu-4) \theta \over 2\nu-4} \cos \theta
 	- {\sin (2\nu-3) \theta \over 2\nu-3} \right\}
\Biggr ],
\end{eqnarray}
and $\nu=3/4 + (\mu + n)/2$.  To normalize the $K$-factor, we also need
the overlap between the Feynman amplitude and the wave function
without the final-state interaction.  By using the wave function
$\psi_0 (r) = {\sin p r}/{p r}$ for the S-state without the Coulomb
potential, we obtained the amplitude without the
final-state interaction as given by
\begin{eqnarray}
\label{eq:down}
 \langle{\Psi}_0 | \Phi_{ab} \rangle
	= {\sqrt{2} e^2 \over m(E+m)} 
	  {\bf k} \cdot ( {\bbox{\epsilon}}_1 \times {\bbox{\epsilon}}_2 )
	{\cal B},
\end{eqnarray}
where the factor $\cal B$ is
\begin{eqnarray}
{\cal B}={(E+m)^2 + \epsilon_w^2 -m_w^2 \over kp}
	 {\cal I}{\rm m} 
	 \left\{ ( \cot \theta^* -{2m \over k})(\theta^* \cot \theta^* -1) 
	  +\theta^* \right\},
\end{eqnarray}
and $\theta^*$ is a complex conjugate of $\theta$.  Then the ratio
between the absolute squares of Eqs.(\ref{eq:up}) and (\ref{eq:down})
is the relativistic expression of the $K$-factor,
\begin{eqnarray}
\label{eq:kfinal}
K &=& \left| 
	{\Gamma(a) \over \Gamma(b)}
	e^{\pi \eta /2} 
	{{\cal A} \over {\cal B}} \right |^2.
\end{eqnarray}
We can identify the factor $|\Gamma(a)e^{\pi \eta /2} / \Gamma(b)|^2$
as closely related to the Gamow factor $G(\eta)$.  One can show that
\begin{eqnarray}
\left|{\Gamma(a) \over \Gamma(b)}
        e^{\pi \eta /2} \right |^2 = G(\eta) \left| {\Gamma(\mu+1/2)
\over \Gamma(2\mu+1)}\right|^2 
\prod_{j=0}^{\infty} \left ( 1+{\mu - 1/2 \over
1+j} \right )^2 \left ( 1 + {({3\over 2} + \mu + 2 j) ( {1 \over 2} -
\mu) \over (\mu + {1 \over 2} + j)^2 + \eta^2 } \right ) .
\end{eqnarray} 
Therefore, the proper treatment of the dynamics of the
interacting particles leads to the modification of the Gamow factor
$G(\eta)$ of Eq.\ (\ref{eq:80}) by a factor $\kappa$ given by
\begin{equation} 
K=G(\eta)\kappa 
\end{equation} 
where 
\begin{eqnarray}
\kappa=\left| {\Gamma(\mu+1/2) \over \Gamma(2\mu+1)} \right|^2 
\prod_{j=0}^{\infty} \left ( 1+{\mu - 1/2 \over 1+j} \right )^2 \left
( 1 + {({3\over 2} + \mu + 2 j) ( {1 \over 2} - \mu) \over (\mu + {1
\over 2} + j)^2 + \eta^2 } \right ) \left | {\cal A \over
\cal B} \right|^2 .
\end{eqnarray} 
Here the energy of the fermion is same to the energy of a gluon($E=k$)
in a center of mass frame
and $(E+m)^2 + \epsilon_w^2 -m_w^2 =2k(k+m)$.
In the limit of $\alpha \rightarrow 0$ or $v\rightarrow 0$, the factor
$\kappa$ goes to 1 and the $K$-factor is consistent with the Gamow factor.

\begin{figure}[h]
\includegraphics[scale=0.60]{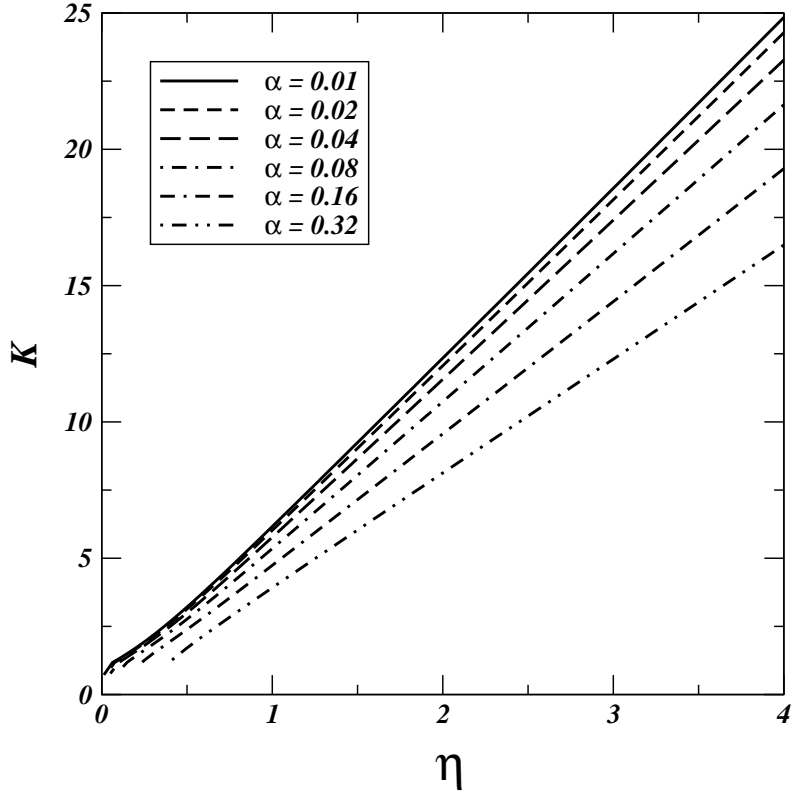}
\caption{The $K$-factor
versus $\eta$ for various values of $\alpha$.}
\end{figure} 
\vspace*{0.1cm}\noindent

We note that the center-of-mass energy $\sqrt{s}$ 
in units of the rest mass of
the produced particle is a function of $\eta/\alpha$:
\begin{eqnarray}
 { \sqrt{s} \over m }   = \sqrt{ 2 \left ( 1 + { {\eta/\alpha} 
\over \sqrt{ \eta^2/\alpha^2 -1}} \right )} .
\end{eqnarray}
Various other kinematic variables, such as ${k / m} = {\sqrt{s} / 2
m}$ and ${p / m} = \sqrt{ {s/ 4m^2 }-1}$ can be similarly expressed as
a function of $\eta/\alpha$.  From these relations and the relation
between the $K$-factor and $\eta$ and $\alpha$, we can find out the
$K$-factor for the production of a pair of particles in a specific
kinematic configuration.

\begin{figure}[h]
\includegraphics[scale=0.60]{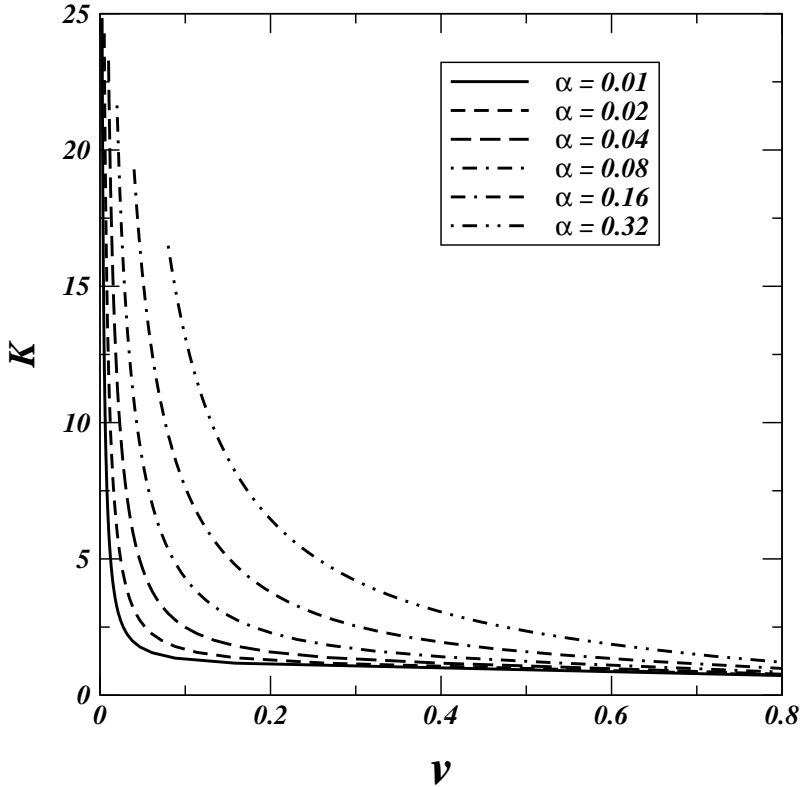}
\caption {The $K$-factor
versus the velocity $v$ for various values of $\alpha$.}
\end{figure} 

We showed the behavior of the $K$-factor as a function of $\eta$ in
Fig. 2 for various values of $\alpha$.  The solid curve gives the
$K$-factor for $\alpha=0.01$ and the dotted curve gives the $K$-factor
for $\alpha=0.32$.  For a fixed value of $\alpha$, the $K$-factor
decreases as $\eta$ decreases.  This is consistent with the
expectation that the effects of the final-state interaction diminish
as the velocity becomes relativistic.  The limiting value is
$K$=1 as $\eta \rightarrow \alpha$(or $v \rightarrow 1$).  
Figure 2 also shows that
for a given value of $\eta$, the $K$-factor decreases as
$\alpha$ increases.  It should be noted that the same value of $\eta$
corresponds to different velocities $v$ for different values of
$\alpha$.  To see the effect of final-state interaction as a function
of $\alpha$ for a fixed value of $v$, we plot in Fig.\ 3 the
$K$-factor as a function of $v$.  As one observes, when the velocity
is fixed, the $K$-factor increases as the coupling constant increases,
indicating a greater effect of the final-state interaction as $\alpha$
increases.  For all values of $\alpha$, the $K$-factor decreases with
$v$ and goes to unity as $v$ approaches 1.  The decrease is very rapid
for small values of $\alpha$.


\begin{figure}[h]
\includegraphics[scale=0.80,angle=270]{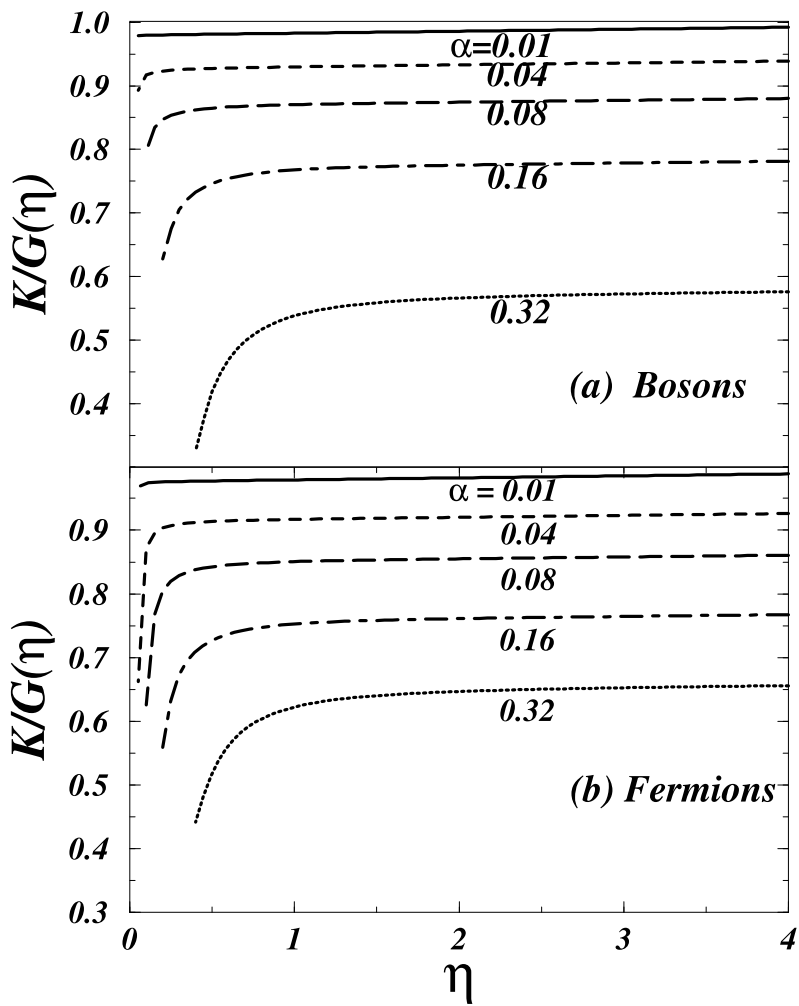}
\caption{The ratio between the $K$-factor and the Gamow
factor for various values of $\alpha$.}
\end{figure}

It is of interest to see how the $K$-factor obtained here is different
from the Gamow factor in non-relativistic physics.  In Fig. 4$(b)$, we
showed the ratio between the $K$-factor and the Gamow factor for
various values of $\alpha$.  As we expect, the ratio is almost 1 for
weak coupling and the use of the Gamow factor is relatively safe
there. However, if we increase $\alpha$ to 0.32, the ratio decreases
significantly.  The Gamow factor over-estimates the magnitude of the
final-state interaction and therefore it cannot be used for the case with strong
coupling.  There is an effective screening of the long-range Coulomb
interaction.  As a consequence, the enhancement due to the long-range
Coulomb-type interaction is reduced.  It can also be observed in Fig.\
4$b$ that the ratio of $K/G(\eta)$ is a relatively slowly varying
function of $\eta$ for $\eta>1$ but drops down rapidly as $\eta$
decreases in the region of small $\eta$ (high $v$).

It is worth pointing out that the expansion of $\cal A$ in Eq.\
(\ref{eq:A1}) is given as a series in powers of
$p/\sqrt{\delta^2+k^2}$ which increases as the velocity $v$ increases.
We still obtain convergent results for $v$ up to about 0.8, but there
is a limit on using such an expansion for greater velocities where
$p/\sqrt{\delta^2+k^2}$ is too large to allow for a convergent
term-by-term summation.  A different expansion method may be needed.
Fortunately, the $K$-factor for this region is so close to unity that
it can be taken to be unity without incurring much error.

We can compare  the results we have obtained for  the case of fermions
with those  for the previous case of bosons.  We  show in
Fig.\  4$a$  the  ratio of  $K/G(\eta)$  for  the final  state
interaction  of two  bosons.  The  $K$ factor  for bosons  is slightly
smaller than  the $K$-factor for fermions and the  difference is greater
for larger values of $\alpha$. This means that the overestimations
by the Gamow factor are larger in bosonic FSI/ISI than in fermionic one.

\section{Conclusions and Discussions}

When a pair of particles are subject to final-state interactions, the
rate of their production is modified.  There will be similar effects if
the particles interact via initial-state interactions.  The
modification is simplest to be taken into account by using the
$K$-factor.  One calculates first the rate for the process when there
were no initial- or final-state interactions, using, for example, the
perturbation theory.  The additional initial- or final-state
interactions are then included by multiplying a $K$-factor as given by
Eq.\ (\ref{eq:rat1}).

For Coulomb-type interactions, the $K$-factor has been traditionally
taken to be the Gamow factor obtained as the absolute square of the
wave function at the origin of the relative coordinate.  With
relativistic Coulomb wave functions, the wave function at the origin
is infinite and the usual method is not applicable.  The $K$-factor
can be obtained as the overlap of the wave function with the Feynman
amplitude.

Our investigation of the $K$-factor for the case of the production of
a pair of scalar particles indicates that there are substantial
deviations from the Gamow factor when the strength of the coupling is
large.  In particular, the proper treatment reduces the magnitude of
the Gamow factor significantly.  The reason for this reduction is that
in the pair production, there is an effective screening of the
Coulomb-type interaction arising from the effective ``exchange'' of
one of the produced particles.

We have presented an explicit formula for the relativistic
modification of the Gamow factor for the production of a pair of
fermions.  Numerical results are also obtained to show the magnitude
of the $K$-factor.  The results of the $K$-factor can be applied to a
class of processes in which the fermion particles are produced and
interacting with a Coulomb-type final-state interaction as for an
example in the production of open charm pairs
\cite{Tai04,Ada04,Adl04}.  Such an application to the production of
heavy quarks systems near the threshold will be of great interest.

\section*{Acknowledgments}
\vspace {-0.5cm}

The authors would like to thank Dr. H. W. Crater for helpful
discussions.  This research was supported by the Korea Research
Foundation under Grant KRF-2001-015-DP0106 and by the Division of
Nuclear Physics, US DOE, under Contract No. DE-AC05-00OR22725 managed
by UT-Battelle, LLC.

\end{document}